\documentclass[structabstract]{aa}  
\usepackage{graphicx}
\usepackage{natbib}
\usepackage{txfonts}
\begin{document}
   \title{Herschel and IRAM-30m Observations of Comet C/2012 S1 (ISON) at 4.5 AU from the Sun\thanks{{\it Herschel} is an ESA space observatory with science instruments provided by European-led Principal Investigator consortia and with important participation from NASA. Based on observations carried out with the 30m telescope,
operated by the Institute de radioastromie millim\'etrique 
(IRAM), which is funded by a partnership of INSU/CNRS (France), MPG
(Germany), and IGN (Spain). 
}} 

   \author{L.\ O'Rourke\inst{1} 
           \and         
           D.\ Bockel\'ee-Morvan\inst{2}
           \and
           N.\ Biver\inst{2} 
           \and
           B.\ Altieri\inst{1}           
           \and
           D.\ Teyssier\inst{1} 
			  \and
           L.\ Jorda\inst{3} 
           \and
		 	  V. \ Debout\inst{2}
           \and
           C. \ Snodgrass\inst{4}
           \and
           M.\ K\"uppers\inst{1}
           \and
           M.\ A'Hearn\inst{5} 
           \and
           T.\ G.\ M\"uller\inst{6}
           \and
           T.\ Farnham\inst{5}
          }

\institute{%
    European Space Astronomy Centre, 
    ESAC, ESA, Villanueva de la Ca\~{n}ada, 28691, Spain;
    \email{lorourke@esa.int}
 \and
	 LESIA, Observatoire de Paris, CNRS, UPMC, 
	 Universit\'e Paris-Diderot, 5 place Jules Janssen, 92195 Meudon, France;
 \and
    Aix Marseille Universit, CNRS, LAM (Laboratoire dAstrophysique de Marseille),
    UMR 7326, 13388 Marseille, France;
 \and
	 Max Planck Institute for Solar System Research, 
	 Max-Planck-Str. 2, 37191 Katlenburg-Lindau, Germany;
 \and
    Dept. of Astronomy, Univ. of Maryland, 
    College Park, MD 20742-2421, USA;
 \and
    Max-Planck-Institut f\"ur extraterrestrische Physik,             
    Giessenbachstra{\ss}e, 85748 Garching, Germany;                                 
    }

   \date{Received; draft}

 
  \abstract
 {The sungrazer comet C/2012 S1 (ISON) (perihelion at $r_h$ = 0.0125 AU from the Sun) was bright and active when discovered in September 2012 at 6.3 AU from the Sun.}  
   {Our goal was to characterize the distant gaseous and dust activity of this comet, inbound, from observations of H$_2$O, CO and the dust coma in the far-infrared and submillimeter domains.}
{We report observations undertaken with the Herschel Space Observatory on 8 \& 13 March 2013 ($r_h$ = 4.54--4.47AU) and with the 30m telescope of Institut de Radioastronomie Millim\'etrique (IRAM) in March and April 2013 ($r_h$ = 4.45--4.18 AU). The HIFI instrument aboard Herschel was used to observe the H$_{2}$O $1_{10}-1_{01}$ line at 557 GHz, whereas images of the dust coma at 70~$\mu$m and 160~$\mu$m were acquired with the PACS instrument. Spectra acquired at the IRAM 30m telescope cover the CO $J$(2--1) line at 230.5 GHz. The spectral observations were analysed with excitation and radiative transfer models. A model of dust thermal emission  taking into account a range of dust sizes is used to analyse the PACS maps.}
{While H$_2$O was not detected in our 8 March 2013 observation, we derive a sensitive 3$\sigma$ upper limit of $Q_{\rm H_2O}$ $<$ 3.5 $\times$ 10$^{26}$ molecules s$^{-1}$ for this date. A marginal 3.2$\sigma$ detection of CO is found, corresponding to a CO production rate of $Q_{\rm CO}$ = 3.5 $\times$ 10$^{27}$ molecules s$^{-1}$. The Herschel PACS measurements show a clear detection of the coma and tail in both the 70~$\mu$m and 160~$\mu$m maps. Under the assumption of a 2-km radius nucleus, we infer dust production rates in the range 10--13 kg s$^{-1}$ or 40--70 kg s$^{-1}$, depending on whether a low or high gaseous activity from the nucleus surface is assumed. 
We constrain the size distribution of the emitted dust by comparing PACS 70 and 160 $\mu$m data, and considering optical data. Size
indices between --4 and --3.6 are suggested.  The morphology of the tail observed on 70~$\mu$m images can be explained by the presence of grains with ages older than 60 days. } 
   {}


  \authorrunning{L.\ O'Rourke et al.}
  \titlerunning{Comet C/2012 S1 (ISON) with Herschel and IRAM at 4.5AU}
   \maketitle
%

\section{Introduction}

Comet C/2012 S1 (ISON) was discovered in September 2012 at $r_h$ = 6.3 AU from the Sun by the Russian astronomers Vitali Nevski and Artyom Novichonok \citep{Nevski2012} using a 15.7-inch reflecting telescope of the International Scientific Optical Network (ISON) near Kislovodsk. Its orbit is nearly parabolic, consistent with a dynamically new comet coming freshly from the Oort cloud. It is peculiar in that it is a sungrazer ($q$ = 0.012 AU  on 28 November 2013) although not of the Kreutz group.

The comet passed approximately 0.072 AU from Mars on 1 October 2013 and passes approximately 0.42 AU from Earth on 26 December 2013. Because of its distant activity and expected exceptional brightness at perihelion, comet C/2012 S1 (ISON)  is the object of a worldwide campaign involving many space and ground-based observatories.

This paper covers observations of comet C/2012 S1 (ISON) undertaken in March and April 2013 ($r_h$ = 4.5--4.2 AU) in the far-infrared with the Herschel Space Observatory \citep{Pilbratt10} using director discretionary time and observations from ground at millimetric wavelengths obtained with the 30m antenna of the Institut de Radioastronomie Millim\'etrique (IRAM).

We first present in Sect.~2 the observations performed with the Heterodyne Instrument for the Far Infrared \citep[HIFI,][]{2010HIFI} of comet C/2012 S1 (ISON) and the data analysis performed on that data set. An upper limit on the water production rate is derived. This is followed in Sect.~3 by an analysis of the IRAM search for CO whereby a marginal detection is described. Sect.~4 presents images of the dust coma obtained with the Photodetector Array Camera and Spectrometer \citep[PACS,][]{Poglitsch10}, which serve to derive the dust production rate and constrain its size distribution. Sect.~5 outlines the conclusions we derive from the work performed in the paper.



\begin{table*}
\caption[]{C/2012~S1 (ISON) H$_2$O and CO observations.}\label{tabobs}
\begin{center}
\begin{tabular}{lcclccccccc}
\hline \noalign{\smallskip}
\multicolumn{2}{c}{line}   &  $\nu$  & Date (UT) & $r_h$$^a$ & $\Delta$$^b$ & $\phi$$^c$ & Integ.  & Line Area$^d$ & \multicolumn{2}{c}{Production rate  (mol. s$^{-1}$)} \\
          &                &  (GHz)  & yyyy/mm/dd.dd  &  (AU) &  (AU) &  ($^{\circ}$) & time (min) & (mK km s$^{-1}$)      &   $T_{\rm kin} = 8$ K     & $T_{\rm kin} = 20$ K \\
\hline \noalign{\smallskip}
H$_2$O  & $1_{10}$--$1_{01}$ & 556.936 & 2013/03/08.13   & 4.54 & 4.07 & 11.8 &     177 & $ < 5.4^e$     & $ < 4.8\times10^{26}$$^f$ & $ < 3.5\times10^{26}$$^f$ \\
 CO     & 2--1             & 230.538 & 2013/03/15.70    & 4.45 & 4.10 & 12.5 &      21 & $40\pm18$    & $4.5\pm2.0\times10^{27} $ & $3.8\pm1.7\times10^{27} $ \\
 CO     & 2--1             & 230.538 & 2013/04/06.66--08.67 & 4.18 & 4.22 & 13.7 &   47 & $32\pm13$    & $3.5\pm1.4\times10^{27} $ & $2.9\pm1.2\times10^{27} $ \\
 CO     & 2--1             & 230.538 & 2013/03/15.7--04/08.7 & 4.27 & 4.19 & 13.0 & 68 & $38\pm12$    & $4.2\pm1.3\times10^{27} $ & $3.5\pm1.1\times10^{27} $ \\
\hline
\end{tabular}
\end{center}
{\bf Notes.}  $^{(a)}$ $r_h$ is the comet's heliocentric distance at the time of observation.
 $^{(b)}$ $\Delta$ is the distance from target to the observer. 
 $^{(c)}$ Phase Angle.
 $^{(d)}$ Line area in the main beam brightness temperature scale computed in a ($-0.6$,$+0.6$) km s$^{-1}$ window.
$^{(e)}$ From the HRS spectrum, the upper limit from the WBS spectrum is 4.2 mK km s$^{-1}$.
$^{(f)}$ The values provided for the gas production rates are 3$\sigma$ upper limits.
\end{table*}

\section{Herschel HIFI observations search for H$_2$O}
\label{sec:hs}

Comet C/2012 S1 (ISON) was observed with the HIFI instrument on Herschel on 8 March 2013 UT (ObsId: 1342266412) with a total on-target integration time of 2.9 h and a line of sight velocity (delta-dot) of 8.24 km s$^{-1}$. Further geometric circumstances are presented in Table 1.

We searched in the upper sideband of the HIFI band 1a mixer for line emission from the fundamental {\it ortho}-H$_2$O $1_{10}$--$1_{01}$ line at 556.936 GHz. Our observation was performed using both the Wide Band Spectrometer (WBS) and the High Resolution Spectrometer (HRS). The spectral resolution of the WBS is 1 MHz (0.54 km s$^{-1}$ at the frequency of the observed line), while the HRS was used in its high-resolution mode with a resolution of 120 kHz (0.065 km s$^{-1}$). The observing mode used was the frequency-switching observing mode with a frequency throw of 94.5 MHz and without a reference position on the sky.
While the advantage of using this mode is that the on-target integration time is maximized, the downside is that the statistical noise can be underestimated due to strong baseline ripples and uncertainties in their removal.

\begin{figure}
\centering
\resizebox{9cm}{!}{\includegraphics{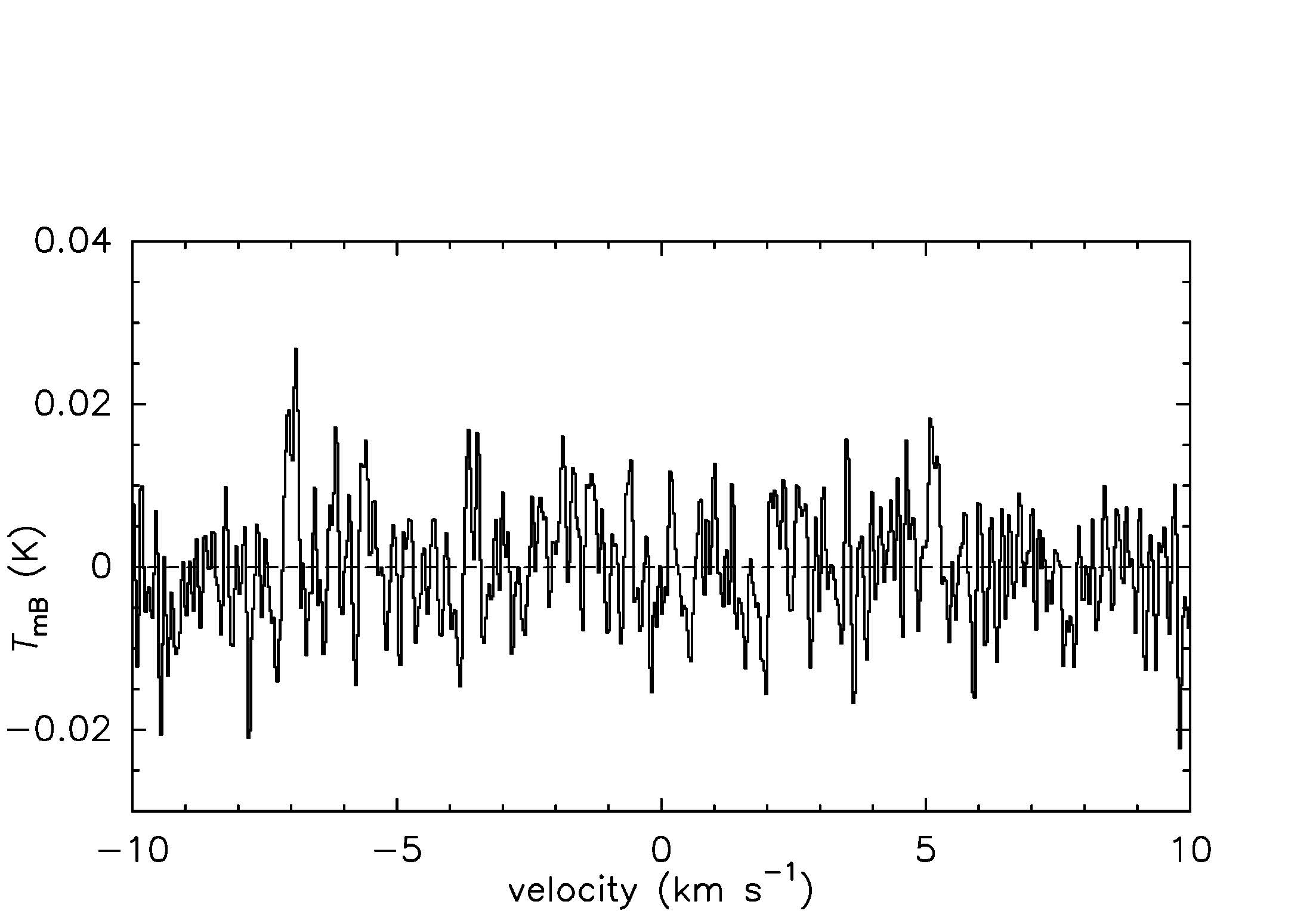}}
\caption{Averaged spectrum of the two orthogonal polarisations of the H$_{2}$O line $1_{10}$--$1_{01}$ at 556.936 GHz towards comet C/2012~S1 (ISON) obtained by Herschel/HIFI on 8.1 March 2013 UT with the HRS spectrometer. 
The vertical scale is the main-beam brightness temperature. 
The horizontal scale is the Doppler velocity in the comet rest frame. }
\label{fig:h2o}
\end{figure}

Initial processing of the HIFI data set was performed using the standard HIFI processing pipeline of HIPE v9.2 (Herschel interactive processing environment, \citep{Ott2010}). This processing served to reduce the data to calibrated level-2 data products.
The main beam brightness temperature scale was computed using a forward efficiency of 0.96 and beam efficiency of 0.75. Vertical and horizontal polarizations were averaged weighted by the mean square amplitude in order to increase the signal-to-noise ratio. The pointing offset of both the vertical and the horizontal polarization spectra is 3.3\arcsec~in band 1a when compared to the target position. The Half Power Beam Width (HPBW) is 38.1$\arcsec$ at 557 GHz. No hint of an H$_{2}$O line is found in the resultant spectra. We show the HRS averaged spectrum of the two orthogonal polarisations in Fig.~\ref{fig:h2o}.

To derive an upper limit on the water production rate, we used a molecular excitation model which calculates the population of the rotational levels of water as a function of nucleocentric distance.  The excitation model includes collisions of water and electrons in the inner coma, infrared pumping of the vibrational bands, and treats self-absorption using the escape probability formalism \citep[e.g.,][]{biv07,zakharov2007}. Model calculations were performed with input parameters which include the electron density, the gas expansion velocity, and the gas kinetic temperature $T_{\rm kin}$ (used to control the molecular excitation in the collisional region). Synthetic spectra were computed considering the transfer of line radiation of ortho-water in the cometary atmosphere. Since the electron density in the coma is not well constrained, an electron density scaling factor of $x_{n_{e}}$ = 0.2 with respect to the standard profile derived from observations of comet 1P/Halley has been used \citep[e.g.,][]{hart10}. For the radial gas density profile, we adopted the standard spherically symmetric Haser distribution. We assumed an expansion velocity of 0.35 km s$^{-1}$ derived from the CO $J$(2--1) line profile (Sect.~\ref{sec:CO}) following similar techniques as described in, e.g. \citep{biver2002} and \citep{ORourke13}. The upper limit on the water production rate was calculated against two different values of the kinetic temperature, 8 K and 20 K. The 8K kinetic temperature value is inferred from that derived for comet 29P/Schwassmann-Wachmann 1 \citep{Gunnarsson2008} and the 20K temperature value from that derived for comet Hale-Bopp \citep{biver2002} when both were at the same heliocentric distance as ISON at the time of the observations.

Table~\ref{tabobs} provides the 3$\sigma$ upper limit on the line integrated intensity and the corresponding production rate. For $T_{\rm kin}$ = 20 K, we derive a sensitive 3$\sigma$ upper limit of $Q_{\rm H_2O}$ $<$ 3.5 $\times$ 10$^{26}$ molecules s$^{-1}$.

\section{IRAM observations of CO}
\label{sec:CO}

\begin{figure}
\centering
\resizebox{9cm}{!}{\includegraphics{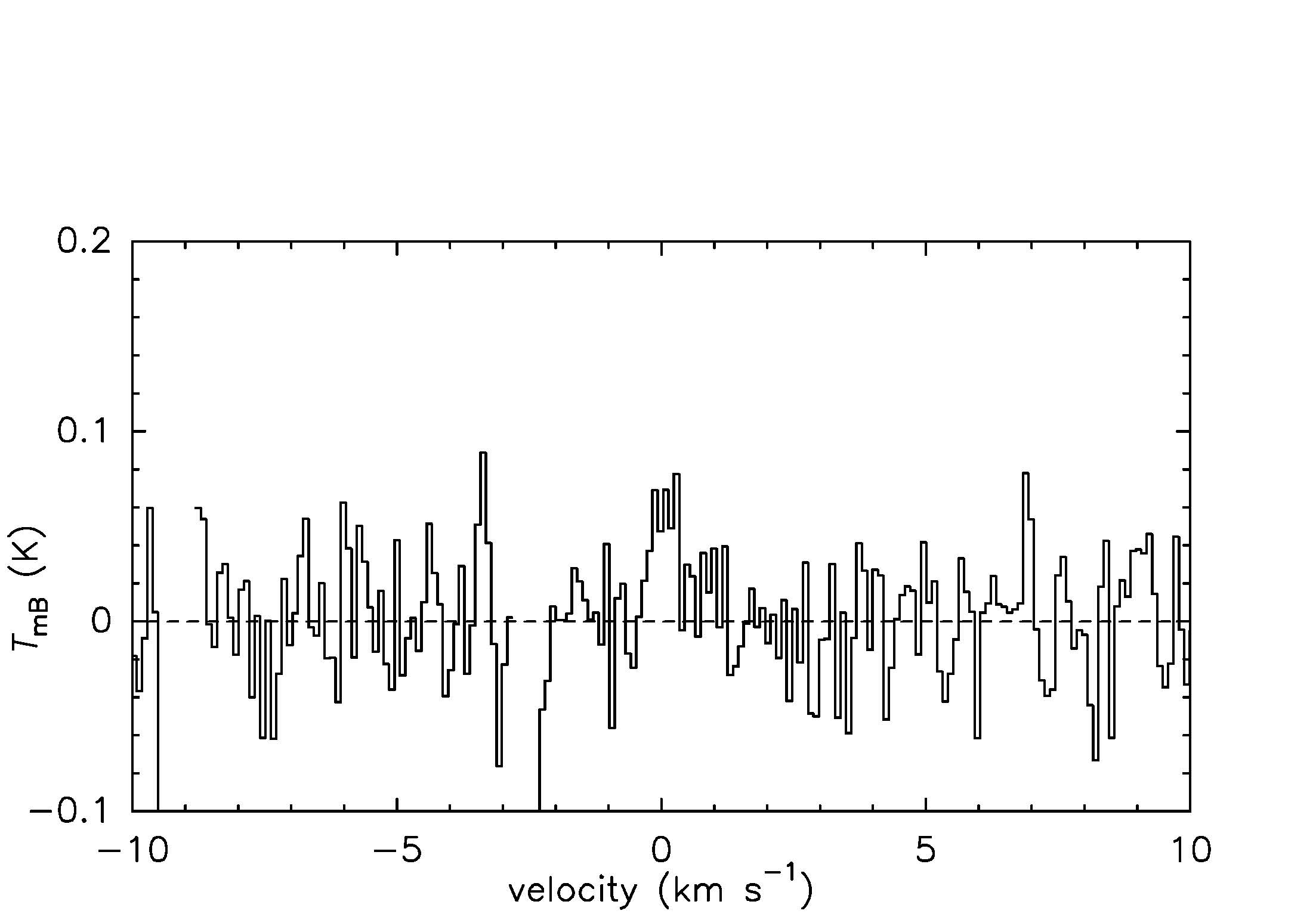}}
\caption{Average CO $J$(2--1) spectrum obtained with the IRAM 30m radio telescope
between 15.7 March and 8.7 April 2013, using the high resolution Vespa autocorrelator.
The vertical scale is the main-beam brightness temperature. 
The horizontal scale is the Doppler velocity in the comet rest frame. }
\label{figco}
\end{figure}

Comet C/2012~S1 (ISON) was observed with the IRAM 30m radio telescope on 15 March, 6 and 8 April 2013 UT (See Table~\ref{tabobs} for the geometric circumstances of the observations). These were short observations inserted in the pre-existing schedule. Observations were done using the wobbling secondary mirror in March, and in position switching mode in April with a reference sky at 3\arcmin~from the comet. The EMIR 230 GHz receiver was used, and spectra covering 224.9--232.6 GHz and 240.6--248.3 GHz were obtained with the FTS spectrometer at a resolution of 200~kHz. In addition, the Vespa autocorrelator covered the CO $J$(2--1) line at 230.538~GHz with a resolution of 40~kHz (52 m s$^{-1}$, Fig.~\ref{figco}). The IRAM beam size is 10.7$\arcsec$ (HPBW) at 230 GHz. 

A marginal narrow signal was seen at the comet velocity in both periods and with both spectrometers, but the detection level in the average spectrum does not exceed 3--4$\sigma$ (Fig.~\ref{figco}). No other line (e.g., from CS, H$_2$CO, CH$_3$OH) was seen.

The line width suggests a rather low expansion velocity (0.35 km s$^{-1}$) \citet{biver2002}, which is used to derive the CO production rate. The line integrated intensity and corresponding production rate for $T_{\rm kin}$ = 8 and 20 K are given in Table~\ref{tabobs}. For $T_{\rm kin}$ = 20 K, we derive a CO production rate $Q_{\rm CO}$ = 3.5 $\times$ 10$^{27}$ molecules s$^{-1}$. The value derived for $T_{\rm kin}$ = 8 K is of the same order.

\section{Herschel PACS data}

\subsection{Observations} 

Comet C/2012 S1 (ISON) was observed by the Herschel PACS instrument on 13 March 2013 UT with a total on-target integration time of 1.9 h. The geometric circumstances are presented in Table 2.

\begin{figure*}
\centering
\resizebox{14.5cm}{!}{\includegraphics{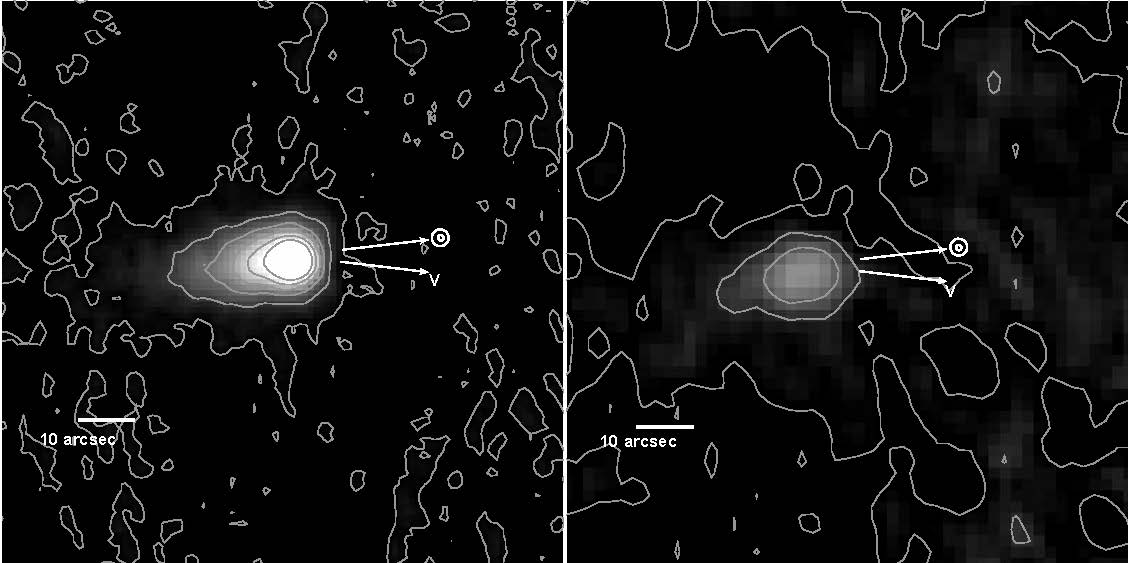}}
\caption{70\,$\mu$m (left, blue map) and 160\,$\mu$m (right, red map) images of comet C/2012 S1 (ISON) observed with PACS on 13 March 2013 UT. The pixel size is 1\arcsec~for the blue map and 2\arcsec~for the red map. Contour levels are stepped by 0.002 in Log, up to 99\% of maximum intensity. North is up and East to the left. The anti Sun direction is at $PA$ = 96.6$^{\circ}$. The arrows indicate the sun direction and projected comet velocity vector. The size of the blue and red images is $\sim$100\arcsec.}
\label{fig:PACS}
\end{figure*}

The observations were performed in the 70-$\mu$m/160-$\mu$m filter combination (so-called blue/red bands). Two observation pairs (ObsId: 1342267433--34; 1342267443--44) were performed (2 scan directions, 70$^{\circ}$ and 110$^{\circ}$) separated by approximately 1.5 h to ensure that the comet had moved during the course of the full observation, supporting easier extraction of the comet signal from the background. Such a strategy was applied in support of data processing of Herschel observations of trans-Neptunian objects (TNOs) \citep[e.g.,][]{kiss13}. We selected six repetitions in each of the two scan-directions for a better characterisation of the background. The PACS measurements were processed using HIPE v9.2 \citep{Ott2010}. Further processing was then performed to merge the four ObsIds for each of the two filters and correct for the motion of the comet during the course of the observation. We used a highpass filter of 50 frames/readout at both wavelengths to remove 1$/f$ noise. Figure~\ref{fig:PACS} provides the two resulting blue and red contour maps.

Using the "Annular Sky Aperture" task of HIPE, the fluxes within a 25\arcsec-radius aperture centred on maximum brightness were extracted for both maps, with the background being subtracted via the same task. The derived fluxes were colour corrected (1.0 at 70\,$\mu$m and 1.06 at 160\,$\mu$m) \citep{Muller11} to obtain monochromatic flux densities at the PACS reference wavelengths. Table~\ref{tabobsp} provides the Herschel PACS observation dataset and accompanying observation information including the final derived total fluxes. We also provide the flux density measured at the photocentre of the images.

\begin{table*}
\caption[]{PACS observations of comet C/2012~S1 (ISON).}\label{tabobsp}
\begin{center}
\begin{tabular}{lcclccccccc}
\hline \noalign{\smallskip}
Wavelength   &  Date (UT) & $r_h$ & $\Delta$ & $\phi$ & Integ.  & Total flux$^a$ & Flux at photocentre$^{a,b}$ \\
      ($\mu$m)    &  (yyyy/mm/dd.dd)  &  (AU) &  (AU) & ($^{\circ}$) &time (min) & (mJy)      &   (mJy/pxl) \\
\hline \noalign{\smallskip}
70  & 2013/03/13.8  & 4.47 & 4.10 & 12.5 & 56.5 & 188 $\pm$ 9 & 1.48 $\pm$ 0.07\\
160  & 2013/03/13.8 & 4.47 & 4.10 & 12.5 & 56.5 & 51 $\pm$ 3 & 0.72 $\pm$ 0.04\\
\hline
\end{tabular}
\end{center}
{\bf Notes.} $^{(a)}$  Colour-corrected flux in an aperture of 25\arcsec~radius, with the sky reference between 40 and 50\arcsec. The quoted uncertainty is 5\% corresponding to the uncertainty in the flux calibration.
$^{(b)}$  The pixel size is 1 and 2\arcsec, at 70 and 160 $\mu$m, respectively.
\end{table*}

\subsection{Dust coma and tail morphology}
\label{sec:morphology}

In both the blue and red maps, there is clear extended emission -- although less so in the red 160-$\mu$m map. Original assumptions that the red map might provide an estimate on the object size were discarded based upon our use of a Thermophysical Model \citep{ORourke12,Muller12}. Indeed, the radius of comet ISON's nucleus has been measured to be less than $R_N$ = 2 km from measurements with the Hubble Space Telescope \citep{Li2013}. Assuming a low albedo (0.03) inert object with a maximum radius of 2 km, with the same observing conditions (distance, phase angle) as PACS, the expected nucleus flux is about 1.5--2 mJy at 70\,$\mu$m, i.e., well below the observed signal (Table~\ref{tabobsp}).  At 160\,$\mu$m the object emission would be below 0.5 mJy approximatively. Effectively, in both cases, the nucleus signal falls within the noise, and the detected signal is from emitted dust. In that respect, we exclude any detection of the nucleus in our observations and focus on the flux measured being primarily from the dust coma and the tail.

At the time of the PACS observations, the position angle of the extended Sun-target radius vector
was 96.6$^{\circ}$, and the negative of the heliocentric velocity vector, as seen in the observer's plane-of-sky, 
was 83.2$^{\circ}$. The tail direction in the blue and red images is along the Sun-comet vector (Fig.~\ref{fig:PACS}), 
as observed on optical images \citep{Meech2013}. The coma is clearly less extended in the 160 $\mu$m image (Fig.~\ref{fig:PACS}). 

We studied the mean radial profile of inner dust-coma emission, performing azimuthal averages. The width of the radial profile in the 70-$\mu$m map is 9.6$\arcsec$, consistent with the expected width for an 1/$\rho$ brightness distribution convolved with the PSF \citep[half width at half maximum of 5.5$\arcsec$,][]{Lutz2012}, where $\rho$ is the distance to the nucleus. For the 160-$\mu$m map, the width is $\sim$ 12$\arcsec$, i.e., on the order of the PSF at this wavelength (11$\arcsec$), whereas a width of 17$\arcsec$ is expected for steady state dust production, not considering dynamics effects \citep{DBM2010}. This suggests an excess of slow large particles in the 160-$\mu$m image. With the expected terminal velocity of 100-$\mu$m dust particles being typically 10 m s$^{-1}$ using our dust models described below, we estimate that the travel time of these particles through the PSF is $\sim$ 15 days.

\begin{figure*}
\centering
\begin{tabular}{cc}
\includegraphics[width=6.4cm]{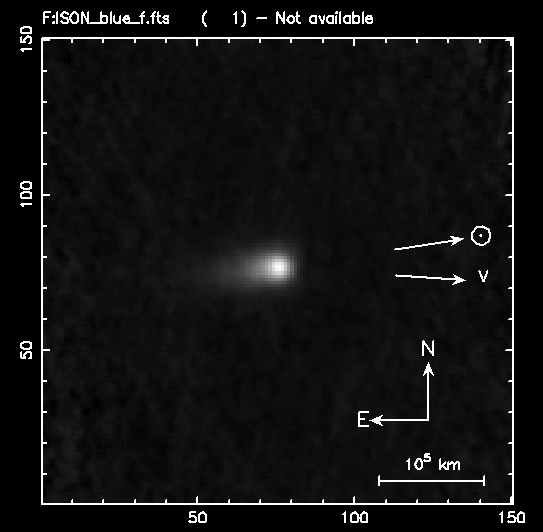}
\includegraphics[width=6.4cm]{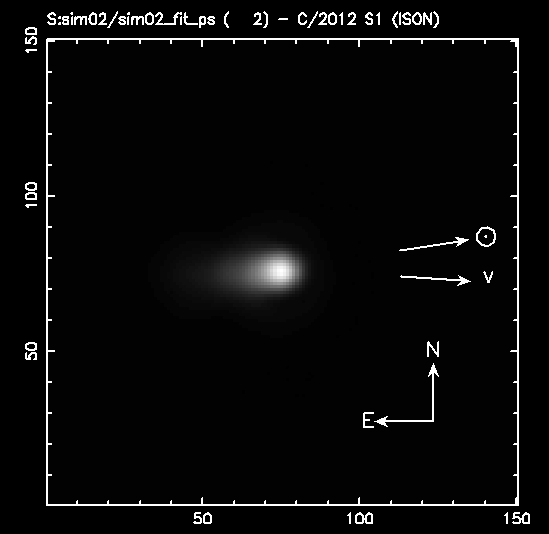}
\end{tabular}
\caption{
A comparison of the dust tail of comet C/2012 S1 (ISON) observed in the blue 
channel of PACS (left image) with a dynamical model of the dust tail
(right image). For details of the model, see text.
The direction of the Sun and that of the projected velocity vector of the comet
are indicated with arrows. The axes units are in arcseconds.}
\label{fig:imgfit}
\end{figure*}

We performed a dynamical simulation of the coma of comet C/2012 S1 (ISON) to explain the structure
of the dust tail. We used a Monte-Carlo program which
computes the trajectory of dust grains under reduced gravity, following a
Finson--Probstein approach.
The relationship between the $\beta$ coefficient and
the radius of the grains has been calculated assuming spherical olivine dust
particles (MgFeSiO$_4$) and Mie theory, where $\beta$ is equal to the ratio of the accelerations
due to radiation pressure and solar gravity \citep{Burns}. 
We assume hereafter that particles with a diameter $2\pi\ a < \lambda$ do not 
contribute significantly to the infrared signal because their absorption 
efficiency factor $Q_{abs}(a,\lambda)$ drops rapidly below this limit 
\citep{Hulst1981}. 
We therefore consider a distribution of dust particles with radii 
$a = 11-100\ \mu\mathrm{m}$, velocities 
$v_d = 0.04-4\ \mathrm{m}\ \mathrm{s}^{-1}$ and age
$\tau_d < 360\ \mathrm{days}$.
The size of the test particles corresponds to $\beta = 0.0015-0.015$.
The maximum terminal dust velocity of $4\ \mathrm{m}\ \mathrm{s}^{-1}$ has been calculated using a ``trial and test'' process.
We used synthetic images of the dust tail for a wide range of size \&
velocity, which we compared visually to the PACS image to 
determine what dust velocities could reproduce both the extension of the 
dust coma in the solar direction and the width of the dust tail in the 
anti-solar direction. The minimum terminal velocity has been chosen 
arbitrarily, and we could not constrain its value from our simulations.
The output of the simulation is a series of 1000 images, in units of 
number of grains,
for 10 bins of sizes, dust velocities and delays \citep[see][]{Jorda2007}.
The images are combined linearly to reproduce the spatial distribution of dust
observed in the blue PACS image.
The results of the comparison are illustrated in Fig.~\ref{fig:imgfit} (right 
panel), which shows our model of the dust tail.

The tail is observed to have a projected length of up to $70\,000\ \mathrm{km}$
in the anti-solar direction (see Fig.~\ref{fig:imgfit}, left panel).
Our simulation shows 
that grains with ages $\tau \geq 60\ \mathrm{days}$ are required to 
reproduce the observed length of the dust tail.
This is explained by the weak radiation pressure at such large distances 
from the Sun combined with the low values of the $\beta$ parameter.
In our simulation, the unresolved central pixels contain grains of ages 
equal to $2-20\ \mathrm{days}$, depending on their terminal velocities consistent with our analysis of the red image above.
The grains described above would be the large dust distribution
counterpart of the smaller micron-size dust observed in the optical tail by 
\cite{Meech2013} between February and May 2013.

\subsection{Dust production rates and size distribution from PACS data}

Analysis of the ISON PACS images to determine the dust production rate $Q_{\rm
dust}$ was performed by comparing the flux density measured on the brightest pixels of the blue and red maps (Table~\ref{tabobsp}) with that of a model of dust thermal emission. 

The model used for this study is the same as that applied to an equivalent PACS data set of comet C/2006 W3 (Christensen) \citep{DBM2010}, and of the centaurs (2060) Chiron and (10199) Chariklo \citep{Fornasier2013}. The basic principles of this model are given in \citet{jew90}. Absorption cross-sections calculated with the Mie theory are used to compute both the temperature of the grains,
solving the equation of radiative equilibrium, and their thermal emission. Complex refractive indices of amorphous carbon and olivine (Mg:Fe = 50:50) \citep{edo83,dors95} are taken as broadly representative of cometary dust. 

We consider a
differential dust production $Q_{\rm dust}$($a$) as a function of
grain radius $a$, described by the size index $\alpha$. The
size-dependent grain velocities $v_{\rm d}$($a$), as well as the
maximum grain radius $a_{max}$, are computed following \citet{cri97}. We assume a nucleus and dust density of 500 kg
m$^{-3}$ \citep{2013Icar..222..550T, 2007Icar..187..306D, 2007Icar..191S.176R, 2012ApJ...744...18N}. Both the maximum grain size and dust velocities 
critically depend on the nucleus size and the gas production rate at the surface, which
are both poorly constrained. For the nucleus size, as a nominal value, our model assumes a nucleus radius of $R_N$ = 2 km, consistent with the size limit inferred by \citet{Li2013}.
For the gas production rate, we utilise as input our results from the HIFI and IRAM measurements. 

Recognising that the HIFI inferred water production rate is at least one order of magnitude weaker than the CO production rate (Table~\ref{tabobs}), we assume a CO-dominated coma with a CO production rate of 4 $\times$ 10$^{27}$ molecules s$^{-1}$, corresponding to the value measured at IRAM on 15 March 2013 (Table~\ref{tabobs}). However, we would normally expect the CO$_2$ production rate to compete with or even dominate the CO production, based on measurements of the CO and CO$_2$ production rates in other comets at large heliocentric distances \citep{Crovisier1997,biver2002,Ootsubo2012}. With no measurement of the CO$_2$ production rate existing for comet ISON in the March 2013 timeframe, and the IRAM marginal detection not confirmed by observations undertaken at the James Clerk Maxwell Telescope \citep{Meech2013}, we look to {\it Spitzer} observations from the June 2013 timeframe to contribute to our analysis.

\citet{Lisse2013} suggest a CO$_2$ production rate of 1.9 $\times$ 10$^{26}$ molecules s$^{-1}$ for 13 June (r$_h$ = 3.34 AU), well below the CO production rate measured on 15 March 2013. \citet{Meech2013} suggested that the activity of comet ISON was CO$_2$ dominated at large heliocentric distances inbound but experienced a long slow outburst between 6--3.5 AU triggered by the sublimation of CO from deep layers. One cannot rule out however that the high CO production rate we measured in March 2013, if real, might be related to sublimation from icy grains exposing a large cross-section to Sun radiation. 

For completeness therefore, and based upon the above analysis, we have also considered a gas production rate from the surface ten times lower than measured at IRAM i.e., equal to 4 $\times$ 10$^{26}$ molecules s$^{-1}$. For comparison, this value is slightly larger than the value of 2.5 $\times$ 10$^{26}$ molecules s$^{-1}$ deduced from the correlation between the heliocentric magnitude and the CO production rate observed in comet Hale-Bopp \citep{Biver2001}. Note that we assume a CO-dominated activity, but similar results are obtained for the dust production rate using CO$_2$ instead.

\citet{Meech2013} conducted Finson-Probstein modelling (\citet{Finson1968}) of the dust tail of comet ISON, and deduced a dust expansion speed of 10 m s$^{-1}$ assuming that micron-sized dust grains dominated the optical appearance of the comet. Such a low velocity would be more consistent with a low surface activity (our dust model predicts a velocity of 35 m s$^{-1}$ for 1--$\mu$m radius particles in the low activity case, Table 3). 
Similarly, our Finson-Probstein modelling of the PACS 70$\mu$m image suggests low grain velocities, though these velocities pertain to 
old grains emitted when the comet activity was lower (Sect.~\ref{sec:morphology}).
       
Finally, we assumed that the dust and gas production is isotropic and that the local density of the individual dust particles follows a $1/r^{2}$ law, where $r$ is the cometocentric distance. 

Table~\ref{tab:qdust} provides the dust velocity and maximum size computed by our model, and the dust production rates that match the the flux density measured on the brightest pixel of the 70 $\mu$m and 160 $\mu$m maps. We described the PSF by a gaussian function
of width 5.5\arcsec and 11\arcsec~at 70 $\mu$m and 160 $\mu$m, respectively \citep{Lutz2012}. We considered a size index between --4 and --3, which corresponds to the range of size indices measured in cometary comae. 

For olivine grains, dust production rates derived from the 70 $\mu$m and 160 $\mu$m maps are consistent with each other within 30\% in the range of size index considered, with a best fit obtained for $\alpha$ $\sim$ --3.65 to --3.6. On the other hand, for carbon grains, the best match is obtained for a steeper size distribution with $\alpha$ $\sim$ --4. In this case there is a factor of $\sim$ 2 discrepancy between 70 $\mu$m and 160 $\mu$m in the derived $Q_{\rm dust}$ for $\alpha$ = --3.

For $Q_{\rm CO}$ = 4 $\times$ 10$^{27}$  molecules s$^{-1}$ (i.e., 190 kg s$^{-1}$ of CO) and the size index which explains both 70 $\mu$m and 160 $\mu$m data, we infer $Q_{\rm dust}$ $\sim$ 70 kg s$^{-1}$ of olivine grains or $Q_{\rm dust}$ = 40 kg s$^{-1}$ of carbon grains. Dust production rates are lower between 10 (model with carbon) and 13 kg s$^{-1}$ (olivine), when a low gas activity at the surface of $Q_{\rm CO}$ = 4 $\times$ 10$^{26}$ molecules s$^{-1}$ (19 kg s$^{-1}$ of CO) is considered, since the maximum grain size is then lower (Table~\ref{tab:qdust}).

\begin{figure}
\centering
\resizebox{9cm}{!}{\includegraphics{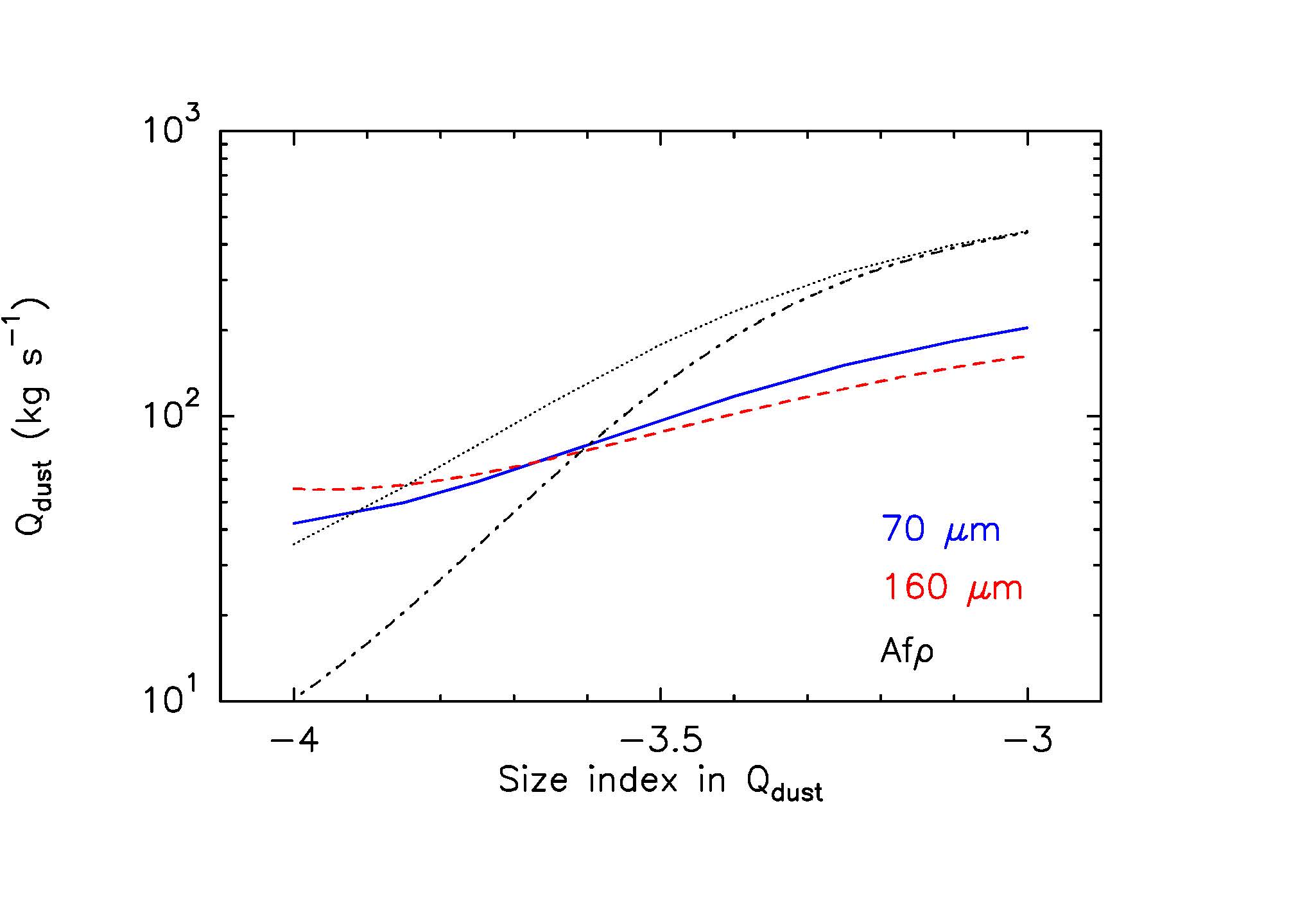}}
\caption{Dust production rates derived from the PACS 70\,$\mu$m (blue curve) and 160\,$\mu$m (red curve) maps of comet C/2012 S1 (ISON) obtained on March 13 2013 UT.  The black curves show values derived from $Af\rho$ = 900 cm. Plain lines are for $a_{\rm min}$ = 0.1 $\mu$m. Dashed lines are for $a_{\rm min}$ = 1 $\mu$m. Refractive indices of olivine are used and we assumed $Q_{\rm CO}$ = 4 $\times$ 10$^{27}$ molecules s$^{-1}$.}
\label{fig:Qdust}
\end{figure}

\subsection{Combining PACS and optical data}

One can also compare the production rates derived from PACS data with the visible light proxy for dust production $Af\rho$, in units of cm \citep{Ahearn84}, where $A$ is the bond albedo of the dust, $f$ is the filling factor of dust in the field-of-view, and $\rho$ is the projected radius of the field-of-view. For steady isotropic dust production expanding at constant velocity, $Af\rho$ is independent of aperture size.

Since visible and far-infrared observations of the dust coma probe particles of different sizes, combining these data may be used to constrain the index of the size distribution. Assuming single-size grains, $Af\rho$ is related to $Q_{\rm dust}$ through \citep{jorda1995} :

\begin{equation}
Q_{\rm dust} =  \frac{2}{3} Af\rho \times \frac{\rho_{\rm d} a
v_{\rm d}}{A_{\rm p}}
\end{equation}

\noindent where $a$ is the grain radius, $\rho_{\rm d}$ is the dust density, and $A_{\rm p}$ is the geometric albedo of the dust. A more general formula can be obtained by considering a dust size 
distribution and a size-dependent dust velocity \citep{Weiler2003}. We used the same dust size and size-dependent velocity distribution as used for the PACS data. 

On 13 March 2013 at 06:00 UT, on the same date as PACS observations, an $Af\rho$ value of $\sim$ 900 cm was measured from R band photometry by the consortium of amateur astronomers CARA (Milani \& Bryssinck, personal communication).
This $Af\rho$ value was computed in an aperture of 5\arcsec~radius (15000 km), approximately of the same size as the PACS pixels. For comparison, \citet{Bode} derived an $Af\rho$ normalized to a phase angle of 0$^{\circ}$ of 1590 cm on 11 March 2013 from V-band magnitudes measured in an aperture of 10 \arcsec~radius. Considering the applied phase function of $\beta(\phi)$ = 0.64 $\beta(0)$ for $\phi$ = 12.7$^{\circ}$ \citep{sch10} and possible short-term variations, the $Af\rho$ value 
inferred by \citet{Bode} is consistent with the value measured by the CARA project. 

In Table~\ref{tab:qdust}, the dust production rates derived from $Af\rho$ are computed for two assumptions on the minimum grain size $a_{\rm min}$ = 0.1 and 1 $\mu$m. Indeed, for steep size distributions, $a_{\rm min}$ affects the results, in contrast to the $Q_{\rm dust}$ determination from PACS data which does not depend much on $a_{\rm min}$.
We assumed the geometric albedo of the grains to be 0.04, consistent with calculations made by \citet{hanner1981} for fluffy particles.

Comparison between the PACS and  $Af\rho$-derived $Q_{\rm dust}$ (Table~\ref{tab:qdust}) 
suggests a steep size distribution with $\alpha$ between --3.5 and --4, depending on the material and minimum size considered. Figure~\ref{fig:Qdust} displays the comparison for olivine grains and the high activity case. We find that the best fit to both PACS 70/160 $\mu$m data and $Af\rho$ data is obtained for olivine grains, and a size index between --3.9 and --3.6 (Fig.~\ref{fig:Qdust}). For these parameters, $Q_{\rm dust}$ is 13 kg s$^{-1}$ or  50--80 kg s$^{-1}$, for the low and high  activity cases, respectively. Similar conclusions are obtained using the refractive indices of carbon grains.

We recognise that the above results are assumption-dependent. For example, if we use a lower nucleus radius or a different grain density, it impacts the inferred dust production rate significantly. In the case of a nucleus radius assumed to be $R_N$ = 1 km, the inferred dust production rate is 20--30 and 65--165 kg s$^{-1}$ for the low and high activity case, respectively, where the low and high values in the ranges refer to calculations for carbon and olivine grains, respectively. As for the comparison between optical and PACS data, it rests upon the albedo of the grains, which might be, e.g., one order of magnitude lower than assumed if the grains are highly porous \citep[][and references therein]{Lacerda2012}.

\begin{table*}[t]
\caption{Dust parameters and production rates for comet C/2012 S1 (ISON) derived
from PACS 70 and 160 $\mu$m data, and from optical data. } \label{tab:qdust}
\begin{center}
\begin{tabular}{clllllll} \hline\\
 $Q_{\rm CO}$ & material & $a_{\rm max}$ & $v_{\rm d}$  &  $\alpha$ & \multicolumn{3}{c}{$Q_{\rm dust}$ (kg s$^{-1}$)} \\
 &&&&&\multicolumn{3}{c}{----------------------------------------} \\
 (molecules s$^{-1}$)  &            & (mm)    & (m s$^{-1}$)$^a$ &           & 70 $\mu$m$^b$ & 160 $\mu$m$^b$ & $Af\rho$$^c$ \\
\hline \hline  \noalign{\smallskip}
 4 $\times$ 10$^{27}$      &    carbon/silicates   & 9.2  & 1.3--167$^d$  & --3 &  210--350$^e$   & 150--170$^e$ & 435\\
 4 $\times$ 10$^{27}$          &    carbon/silicates    &  9.2   & 1.3--167$^d$  & --3.5 & 100--140$^e$   & 80 & 125--176\\
 4 $\times$ 10$^{27}$          &    carbon/silicates    &  9.2    & 1.3--167$^d$  & --4 &  42  &  36--50$^f$ & 10--35 \\
\hline \noalign{\smallskip}
 4 $\times$ 10$^{26}$          &    carbon/silicates    &  0.92     & 1.3--87$^d$  & --3 &  20--35$^e$   & 14--16$^e$ & 45\\
 4 $\times$ 10$^{26}$          &    carbon/silicates    &  0.92     & 1.3--87$^d$  & --3.5 & 14--18$^e$  & 11--12$^f$ & 18--25 \\
 4 $\times$ 10$^{26}$          &    carbon/silicates    &  0.92     & 1.3--87$^d$ & --4 & 12   & 10--14$^f$  & 4--10 \\
 \hline
\end{tabular}
\end{center}
{\bf Notes.} $^{(a)}$ Dust velocities in the size range $a_{\rm
min}$--$a_{\rm max}$, with $a_{\rm min}$ = 0.1 $\mu$m.
$^{(b)}$  Dust production rate derived assuming $a_{\rm
min}$ = 0.1 $\mu$m.
$^{(c)}$  Dust production rate derived assuming $Af\rho$ = 900 cm and $A_p$ = 0.04. The values for $a_{\rm
min}$ = 0.1 and 1 $\mu$m are given when they differ significantly.
$^{(d)}$  The velocity of 1-$\mu$m sized particles is 35 and 85 m s$^{-1}$ for $Q_{\rm CO}$ = 4 $\times$ 10$^{26}$ and 4 $\times$ 10$^{27}$ molecules s$^{-1}$, respectively.
$^{(e)}$ The lower values correspond to olivine and higher values to carbon grains respectively.
$^{(f)}$ The lower values correspond to carbon and higher values to olivine grains respectively.

\end{table*}

\section{Conclusions}

We have presented a detailed analysis of our Herschel Space Observatory and IRAM observations of comet C/2012 S1 (ISON). The observations were performed in  March and April 2013 when the comet was at a distance of approximately 4.5 AU from the sun on the inbound portion of its orbit. 

Our main findings are summarised below :

1. While H$_2$O was not detected by the Herschel HIFI instrument, we derived a sensitive 3$\sigma$ upper limit of $Q_{\rm H_2O}$ $<$ 3.5 $\times$ 10$^{26}$ molecules s$^{-1}$. 

2. We obtained a marginal 3.2$\sigma$ detection of CO with the IRAM 30m telescope, corresponding to a CO production rate of $Q_{\rm CO}$ = 3.5 $\times$ 10$^{27}$ molecules s$^{-1}$.

3. Our Herschel PACS measurements show a clear detection of the coma and tail in both the 70~$\mu$m and 160~$\mu$m maps. 
The length of the tail observed at 70~$\mu$m images can be explained by the presence of grains with ages older than 60 days. The radial profile in the inner coma observed at 160~$\mu$m suggests an excess of slow particles. With an expected terminal velocity of 100-$\mu$m dust particles being typically 10 m s$^{-1}$ (estimated from our models) the travel time of such slow particles through the PSF is $\sim$ 15 days.


4. Under the assumption of a 2-km radius nucleus, we infer dust production rates in the range 10--13 kg s$^{-1}$ or 40--70 kg s$^{-1}$, depending whether a low (4 $\times$ 10$^{26}$ molecules s$^{-1}$) or high (4 $\times$ 10$^{27}$ molecules s$^{-1}$) gaseous activity from the nucleus surface is considered. Size indices between --4 and --3.6 are suggested.

5. A comparison with optical data taken on the same date as our Herschel PACS measurements was performed. We obtained a satisfactory fit to both PACS 70/160 $\mu$m data and CARA $Af\rho$ data in the case of olivine grains whereby we find that a size index between --3.9 and --3.6 best matches the data. For these parameters, $Q_{\rm dust}$ is 13 kg s$^{-1}$ or 50--80 kg s$^{-1}$, for the low and high activity cases, respectively. 

6. In the scenario of weak surface activity ($Q_{\rm CO}$ = 2.5 $\times$ 10$^{26}$ molecules s$^{-1}$, as derived from the magnitude/$Q_{\rm CO}$ correlation established from Hale-Bopp data), we derive a dust-to-gas ratio of $\sim$ 1 for $R_N$ = 2 km, and in the range 2--3 for $R_N$ = 1 km.
Dust-to-gas ratios lower than 1 are inferred in the high activity case.

\begin{acknowledgements}

We wish to thank the Herschel Project Scientist and Time Allocation Committee for awarding 5 hours of Director Discretionary Time for this observation. 

We thank Andy Pollock for his final proof-reading checks.

We thank G. Milani and E. Bryssinck for providing us $Af\rho$ measurements on behalf the 
Cometary ARchive for Afrho (CARA) consortium (http://cara.uai.it/), and R. Ligustri who performed the observation that we used. 

CS has received funding from the European Union Seventh Framework Programme (FP7/2007-2013) under grant agreement no. 268421.

\end{acknowledgements}


\begin{thebibliography}{}
\bibitem[A'Hearn et al, 1984]{Ahearn84} A'Hearn, M. F, Schleicher, D. G., Millis, R. L., Feldman, P. D., \& Thompson, D. T. 1984, \aj, 89, 579
\bibitem[Biver(2001)]{Biver2001} Biver, N.\ 2001, International Comet Quarterly, 23, 85 
\bibitem[Biver et al.(2002)]{biver2002} Biver, N., Bockel{\'e}e-Morvan, D., Colom, P., et al.\ 2002, Earth Moon and Planets, 90, 5 
\bibitem[Biver et al.(2007)]{biv07} Biver, N., Bockel{\'e}e-Morvan, D., Crovisier, J., et al.\ 2007, \planss, 55, 1058
\bibitem[Bockel{\'e}e-Morvan et al.(2010)]{DBM2010} Bockel{\'e}e-Morvan, D., Hartogh, P., Crovisier, J., et al.\ 2010, \aap, 518, L149
\bibitem[Bodewits et al.(2013)]{Bode}
Bodewits, D., Farnham, T., A'Hearn, M.F. 2013, Central Bureau Electronic Telegrams, 3608
\bibitem[Burns et al.(1979)]{Burns}
Burns, J.A., Lamy, P.L., \& Soter, S. 1979, Icarus, 40, 1
\bibitem[Crifo \& Rodionov(1997)]{cri97} Crifo, J.~F., \& Rodionov, A.~V.\ 1997, \icarus, 127, 319
\bibitem[Crovisier et al.(1997)]{Crovisier1997} Crovisier, J., Leech, 
K., Bockel{\'e}e-Morvan, D., et al.\ 1997, Science, 275, 1904 
\bibitem[Davidsson et al.(2007)]{2007Icar..187..306D} Davidsson, B.~J.~R.,
Guti{\'e}rrez, P.~J., \& Rickman, H.\ 2007, \icarus, 187, 306
\bibitem[de Graauw et al.(2010)]{2010HIFI}
de Graauw, T., Helmich, F.~P., Phillips, T.~G., et al. 2010, \aap, 518, L6
\bibitem[Dorschner et al.(1995)]{dors95} Dorschner, J., Begemann, B., Henning, T., Jaeger, C., \& Mutschke, H.\ 1995, \aap, 300, 503
\bibitem[Edoh(1983)]{edo83} Edoh, J. H. 1983, Ph.D. thesis, Univ. Arizona
\bibitem[Finson \& Probstein.(1968)]{Finson1968} Finson, M.L., Probstein, E.F., \ 1968, ApJ, 154, 353
\bibitem[Fornasier et al.(2013)]{Fornasier2013} Fornasier, S., Lellouch, E., M{\"u}ller, T., et al.\ 2013, \aap, 555, A15 
\bibitem[Gunnarsson et al.(2008)]{Gunnarsson2008} Gunnarsson, M., Bockel{\'e}e-Morvan, D., Biver, N.,  Crovisier, J., Rickman, H., \aap, 484, 537
\bibitem[Hartogh et al.(2010)]{hart10} Hartogh, P., Crovisier, J., de Val-Borro, M., et al.\ 2010, \aap, 518, L150
\bibitem[Hanner et al.(1981)]{hanner1981} Hanner, M.S., Giese, R.H, Weiss, K., \& Zerrull, R. \ 1981, \aap, 104, 42
\bibitem[Jorda(1995)]{jorda1995} Jorda, L. 1995, PhD Thesis, University of Paris 7
\bibitem[Jorda et al.(2007)]{Jorda2007} Jorda, L., Lamy, P., Faury, G., et al.\ 2007, \icarus, 187, 208 
\bibitem[Jewitt \& Luu(1990)]{jew90} Jewitt, D., \& Luu, J.\ 1990, ApJ, 365, 738
\bibitem[Kiss et al.(2013)]{kiss13} Kiss, C., M{\"u}ller, T. G., Vilenius, E., et al.\ 2013,  Experimental Astronomy, accepted ({\tt http://arxiv.org/abs/1309.4212})
\bibitem[Lacerda \& Jewitt(2012)]{Lacerda2012} Lacerda, P., \& Jewitt, D. 2012, \apjl, 760, L2
\bibitem[Li et al.(2013)]{Li2013} Li, J.-Y., Weaver, H.~A., 
Kelley, M.~S., et al.\ 2013, Central Bureau Electronic Telegrams, 3496, 1 
\bibitem[Lutz(2012)]{Lutz2012} Lutz, D. 2012, PACS ICC technical note, PICC-ME-TN-033, {\tt 
http://herschel.esac.esa.int/twiki/bin/view/Public/PacsCalibrationWeb?template=viewprint} 
\bibitem[Lisse et al.(2013)]{Lisse2013} Lisse, C.~M., Vervack, 
R.~J., Weaver, H.~A., et al.\ 2013, Central Bureau Electronic Telegrams, 
3591, 2 
\bibitem[Meech et al.(2013)]{Meech2013} Meech, K.J., Yang, B., Ansdell, M., et al. \ 2013, \apj, submitted
\bibitem[M{\"u}ller et al.(2011)]{Muller11} M{\"u}ller, T. G., Okumura, K., Klaas, U.l. \ 2011,PACS technical report PICC-ME-TN-038, v1.0: {\tt 
http://herschel.esac.esa.int/-twiki/-pub/-Public/-PacsCalibrationWeb/-cc/report/v1.pdf} 
\bibitem[M{\"u}ller et al.(2012)]{Muller12} M{\"u}ller, T. G., O'Rourke, L., Barucci, A. M., Pal, A., Kiss, C., et al.\ 2012, \aap, 548, 36 
\bibitem[Niimi et al.(2012)]{2012ApJ...744...18N} Niimi, R., Kadono, T.,
Tsuchiyama, A., et al.\ 2012, \apj, 744, 18
\bibitem[Nevski \& Novichonok(2012)]{Nevski2012} Nevski, V., \& Novichonok, A.\ 2012, IAU Circular 3238 
\bibitem[Ootsubo et al.(2012)]{Ootsubo2012} Ootsubo, T., Kawakita, 
H., Hamada, S., et al.\ 2012, \apj, 752, 15
\bibitem[O'Rourke al.(2012)]{ORourke12} O'Rourke, L., M{\"u}ller, T.,  Valtchanonv, I., Altieri, B., Gonzalez-Garcia, B., et al.\ 2012, \planss, 66, 192 
\bibitem[O'Rourke al.(2013)]{ORourke13} O'Rourke, L., Snodgrass, C., de Val Borro, M., Biver, N., et al.\ 2013, \apj, 774, 13
\bibitem[Ott(2010)]{Ott2010} Ott, S. 2010, ASP Conference Series, 434, 139 
\bibitem[Pilbratt et al.(2010)]{Pilbratt10} Pilbratt, G.~L., Riedinger, J.~R., Passvogel, T., et al.\ 2010, \aap, 518, L1
\bibitem[Poglitsch et al.(2010)]{Poglitsch10} Poglitsch, A., Waelkens, C., Geis, N., et al.\ 2010, \aap, 518, L2 
\bibitem[Richardson et al.(2007)]{2007Icar..191S.176R} Richardson, J.~E.,
Melosh, H.~J., Lisse, C.~M., \& Carcich, B.\ 2007, \icarus, 191, 176
\bibitem[Schleicher(2010)]{sch10} Schleicher, D. 2010, {\tt http://asteroid.lowell.edu/comet/dustphase.html}
\bibitem[van de Hulst(1981)]{Hulst1981}van de Hulst, H.~C. \ 1981, Light scattering by small particles. New York: Dover, 1981  
\bibitem[	Thomas et al.(2013)]{2013Icar..222..550T} Thomas, P.~C., A'Hearn, M.~F., Veverka, J., et al.\ 2013, \icarus, 222, 550
\bibitem[Weiler et al.(2003)]{Weiler2003} Weiler, M., Rauer, H., Knollenberg, J., Jorda, L., \& Helbert, J.\ 2003, \aap, 403, 313 
\bibitem[Zakharov et al.(2007)]{zakharov2007} Zakharov, V., Bockel{\'e}e-Morvan, D.,
Biver, N., Crovisier, J., \& Lecacheux, A.\ 2007, \aap, 473, 303
\end{thebibliography}
\end{document}